\documentclass[prl,twocolumn,superscriptaddress,preprintnumbers,amsmath,amssymb,floatfix]{revtex4} 
\usepackage{graphicx}

\begin{document}

\title{Pressure induced bcc to hcp transition in Fe: Magnetism-driven structure transformation}
\author{S.~Mankovsky} 
\author{S.~Polesya} 
\author{H.~Ebert} 
\affiliation{Dept.~Chemie/Physikalische  Chemie,  Universit\"at  
M\"unchen,  Butenandtstr.   5-13, D-81377 M\"unchen,  Germany}  
\author{W.~Bensch} 
\affiliation{Institut f\"ur Anorganische Chemie,
  Christian-Albrechts-Universit\"at zu Kiel, Max-Eyth-Str. 2, D-24118
  Kiel,~Germany}    
\author{O. Mathon} 
\author{S. Pascarelli} 
\affiliation{European Synchrotron Radiation Facility, 6 Rue Horowitz
  38043 Grenoble, France} 
\author{J.~Min\'{a}r} 
\affiliation{Dept.~Chemie/Physikalische  Chemie,  Universit\"at
  M\"unchen,  Butenandtstr.   5-13, D-81377 M\"unchen,  Germany}   
\begin{abstract}  
The pressure induced bcc to hcp transition in Fe has been investigated 
via ab-initio electronic structure calculations. 
It is found by the disordered local moment (DLM) calculations that the
temperature induced spin fluctuations result in the decrease of the
energy of Burgers type lattice distortions and softening of the 
 transverse  $N$-point $TA_1$ phonon mode with $[\overline{1}10]$
 polarization. As a consequence, spin disorder in an system  leads to
the increase of the amplitude of atomic displacements.
On the other hand, the exchange coupling parameters obtained in our
calculations strongly decrease at large amplitude of lattice
distortions. 
This results in a mutual interrelation of structural and magnetic degrees 
of freedom leading to the instability of the bcc structure under pressure 
at finite temperature.  
\end{abstract}

\maketitle
\section*{I. Introduction}

Since many years considerable effort has been made to investigate the
problem of nucleation and growth of a new phase upon martensitic
transformation. Despite these attempts, there is no clear understanding
so far of the atomic scale mechanism of the bcc-hcp reconstructive
transformation occurring even in non-magnetic materials. In the case of
Fe - considered here - there are convincing arguments, both from
theoretical and experimental side, that magnetism plays a crucial role
for the stability of 
bcc structure, making the problem more complicated \cite{RKC+99, MBI+04,
   ESEB98,LJ09,LPAV12,RR12}.

The bcc-hcp transformation observed in non-magnetic materials has been
extensively discussed in the literature \cite{LM86,Tol05,
  DGT91,DT94,ILS94,CHH88}. The corresponding mechanism 
suggested by Burgers \cite{Bur34} consists in two types of
simultaneous distortions (see Appendix, Fig. \ref{FIG_Burgers} ): 
(i) opposite displacement of adjacent (110) planes along the
$[110]_{bcc}$  direction, described by the parameter $\delta$, associated
with the transverse  $N$-point $TA_1$ phonon mode with
$[\overline{1}10]$ polarization; (ii) shear deformation along the [001]
direction, characterized by the angle 
$\theta$ between the diagonals in (110) plane, which should change from
$109.5^\circ$ in the case of bcc structure to $120^\circ$ in the case of hcp
structure. The shear modulus is determined by the slope of the  $TA_1$
phonon branch $[\xi,\xi,2\xi]$ with $[11\overline{1}]$ polarization.    

Note that even for non-magnetic materials a phenomenological description of
this type of martensitic transformation is not straightforward: 
as pointed out in
the literature \cite{LM86,Tol05,DGT91,DT94,ILS94,CHH88}, the transition
is discontinuous, having large critical displacements and no
group-subgroup relationship between the symmetries of the initial and
final phases. This causes problems for a 
Landau free energy expansion with respect to an order parameter.  
To deal with first-order transitions the phenomenological
Landau theory was extended \cite{LM86,Tol05,SSLA01} using two order
parameters representing shuffle and shear deformations and have been
applied rather successfully to the bcc-hcp phase transitions in Ti
\cite{PHT+91} and Zr \cite{HPT+91},  associated with the softening of
the $N$ point $TA_1$ phonon mode.  According to these theoretical findings,
already small phonon softening (as it takes place in the case 
of Ti and Zr) can be sufficient for a first order transformation
\cite{PHT+91,HPT+91,SSLA01,LJ09}. 

The theoretical approach used for non-magnetic systems
has been applied to Fe,
showing the important role of magnetism.
In contrast to the non-magnetic bcc metals mentioned above,  
no softening of the $N$ point phonon modes have been observed under
pressure neither experimentally 
up to $10$ GPa \cite{MSN67,BAH67,KB00} nor theoretically 
\cite{ESEB98,HCG+02}. 
However, some DFT based theoretical investigations report about the key
role of the shear stress in Fe under pressure for the bcc-hcp transition
\cite {ZMM00,CLOC04}. 
Their role has been investigated by Sanati et al. \cite{SSLA01} in
application to  Ti and Zr showing that the bcc structure in these materials
is completely stable with respect to shear deformation and only the $N$
point phonon mode is responsible for bcc-hcp 
 transformation. 

Ab-initio investigations by Ekman et al. \cite{ESEB98} clearly showed
that the stability of the bcc phase of Fe is due to
magnetic ordering. They have shown that the Burgers type of lattice
distortion results in the transition to the paramagnetic state at a certain
amplitude of the atomic displacements and that way to the instability of
the bcc structure.   
Liu and Johnson  \cite{LJ09} have analyzed the potential-energy surface and
minimum-energy pathway obtained within ab-initio calculations and
have also found that the magnetization collapse during the shuffle-shear 
(Burgers type) deformation leads to the instability of the bcc phase of Fe
under pressure. 

To investigate experimentally the role of magnetism in the bcc-hcp
transformation, Mathon et al. \cite{MBI+04} have performed 
measurements of near edge X-ray absorption (XANES) including a determination
of the X-ray magnetic
circular dichroism (XMCD) for Fe under pressure. The high sensitivity of
XMCD allows very precise measurements of the ordered magnetic moment on
the absorber at the
magnetic phase transition and to observe its correlation with the  
local geometrical structure monitored via XANES.
The pressure dependence of the XMCD and XANES spectra around the
transition pressure suggests that the magnetic transition
slightly  
precedes the structural one. This finding allowed the authors to ascribe
a leading role in the bcc-hcp transformation to the magnetic order
in the system. Nevertheless, this interrelation requires further
clarification, because, in general, the vanishing of magnetism should
lead immediately to the instability of the bcc state and therefore, no
difference in the transition pressure deduced from XANES and XMCD spectra
should be expected. 

Although these results give information on the origin of the pressure induced
instability of bcc Fe and showing the minimal-energy pathway for the
transition, there is still the question how the instability
condition which needs a certain phonon softening under pressure occurs while no softening could be found.  
One possible way has been suggested by Vul and Harmon
\cite{VH93} in their fluctuation-less mechanism for martensitic
transformations triggered by the defects presented in the crystal.
In the present work we analyze the conditions which can result in the
instability in perfect bcc Fe via lattice fluctuations, in particular
the effect of  
temperature induced spin fluctuations for the bcc-hcp transition. 
These investigations in particular give an answer to the question, why the decrease of the XMCD
signal slightly precedes that of the XANES signal corresponding to bcc
structure upon a pressure increase.

\section*{II. Theoretical investigations}

\subsection*{II.A. Details of calculations}

Spin-polarized  electronic structure calculations have been performed
using the spin-polarized KKR (Korringa-Kohn-Rostoker) 
Green's function method \cite{EKM11} in the fully relativistic
approach. The Generalized Gradient Approximation (GGA) for
density functional theory was used with the exchange-correlation
potential due to Perdew, Burke, Ernzerhof (PBE) \cite{PBE96}.
The potential was treated within the
Full Potential (FP) scheme. For the angular momentum
expansion of the Green's function a cutoff of
$l_{max} = 4$ was applied. To treat spin disorder in the system, the
self-consistent coherent potential approximation (CPA) method was
employed. For the calculation of the x-ray absorption coefficient
$\mu^{\vec{q}\lambda}$ the following expression has been used
\begin{equation}\label{MXGFUN}
\mu^{\vec{q}\lambda} (\omega )\propto
        \sum_{i\,{\rm occ}}
        \langle{\Psi_i}\hat{X}_{\vec q\lambda}
         \,\Im G^{+}(E_i+\omega )\,\hat{X}_{\vec q\lambda}^{\times}
        {\Psi_i}\rangle\,\Theta (E_i+\omega -E_{\rm F})
\end{equation}
where $\vec{q}$, $\omega$, and $\lambda$ stand for the wave vector,
frequency, and polarization of the radiation, and $\hat{X}_{\vec
  q\lambda}$ is the electron-photon interaction operator \cite{Ebe96b}.

The finite temperature magnetic properties have been investigated via
Monte Carlo (MC) simulations based on the classical Heisenberg 
model, using a standard Metropolis algorithm \cite{Bin97}.
The exchange coupling parameters $J_{ij}$ for these calculations are
obtained within the approach described by Lichtenstein \cite{LKG84,LKA+87}.

\subsection*{II.B. Total energy calculations \\
           for the bcc-structure}

\begin{figure}
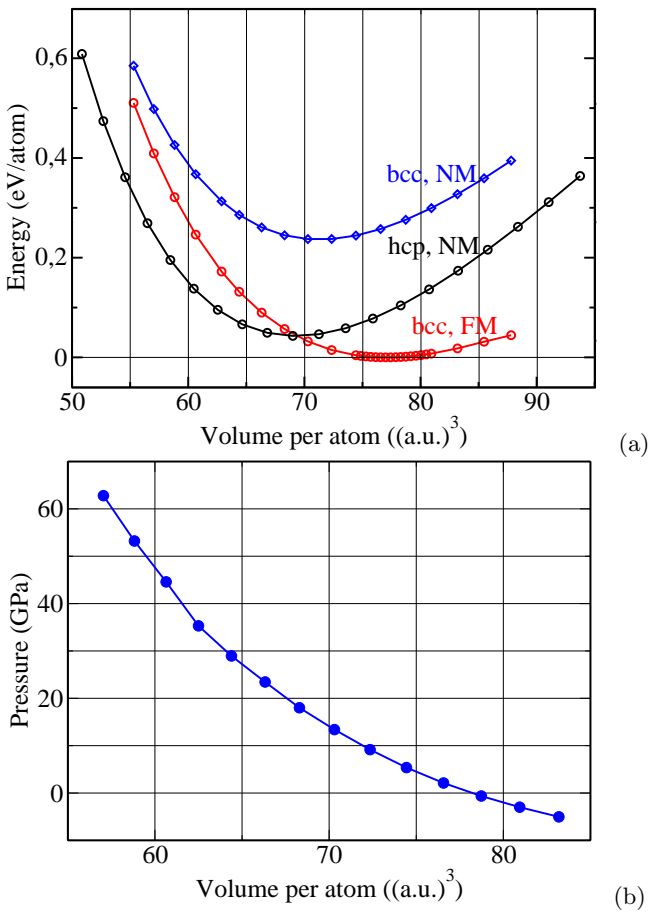

\includegraphics[height=0.33\textwidth,angle=0]{Figure_1_a.eps}\;\;\;(a) 
\includegraphics[height=0.33\textwidth,angle=0]{Figure_1_b.eps}\;\;\;(b)
\caption{\label{FIG_ETOT} (a) Total energy as a
  function of volume for ferromagnetic (FM) and non-magnetic (NM) states
  of bcc-Fe as well as for NM state of hcp-Fe; (b) pressure as a
  function of volume for bcc-Fe.}
\end{figure}

Fig.  \ref{FIG_ETOT} represents the results 
of total energy calculations for
bcc Fe in ferromagnetic (FM) and non-magnetic (NM) states as well as for
hcp structure in the NM state. 
(Below we will use the notation 'NM' and 'PM' to denote the
non-magnetic state with with zero magnetic moments and
the paramagnetic state implying a random distribution of the localized 
magnetic moments, respectively.)  
The calculations for the hcp structure have been performed at a fixed
$c/a$ ratio equal to $1.596$. 
According to the results shown in Fig. \ref{FIG_ETOT}, the minimum of
the total energy corresponds to the equilibrium lattice parameter $a =
5.37$ a.u. in the case of the FM state of bcc-Fe, 
and in the case of hcp-Fe - to $a = 4.66$ a.u. 
The critical pressure evaluated from the equivalence of the enthalpy of the
bcc and hcp phases is equal to 8 GPa in reasonable agreement
with experiment.   

The FM to NM transition leading to the instability of the bcc structure
occurs at the lattice parameter $a \approx 4.6$ 
a.u. ($V \approx 48.7$~(a.u.)$^3$, Fig. \ref{FIG_ETOT}) that corresponds to
a pressure of about 200 GPa. This value is much too 
high when compared directly to experiment. Therefore this type of
instability does not seem to play a role for the real system.

\subsection*{II.C. Total energy calculations \\
        for the distorted bcc Fe}

The energy
of Burgers type lattice distortion in bcc Fe have been calculated
first assuming FM order in the system.  
In contrast to the works of Ekman \cite{ESEB98} and Vul and Harmon
\cite{VH93}, where the authors used two  parameters $\theta$ and $\delta$  to describe the
shear deformation within the (110) plane and
displacements of the neighbor (110) planes with respect to each other
(see Introduction), in the present investigations we consider 
the path from bcc to hcp transformation as suggested by Friak and Sob
\cite{FS08}, which includes both 
deformations using only one parameter $\Delta$ (see
Appendix). This approach avoids the high-energy configurations
occurring upon independent variation of the parameters $\delta$ and
$\theta$, accounting only those being close to the minimum total energy
path \cite{FS08}.
The parameter $\Delta=0$ corresponds to
the bcc structure, while  $\Delta=1$ corresponds to the hcp 
structure  with $c/a = \sqrt{8/3}$.
As soon as additional calculations (not presented here) 
exhibit only a weak dependence on the $c/a$ ratio for the
position of the minimum of total energy as a function of lattice
parameter, the following discussion concerns the results
for $c/a = \sqrt{8/3}$ for the sake of convenience.  

\begin{figure}
\includegraphics[width=0.45\textwidth,angle=0]{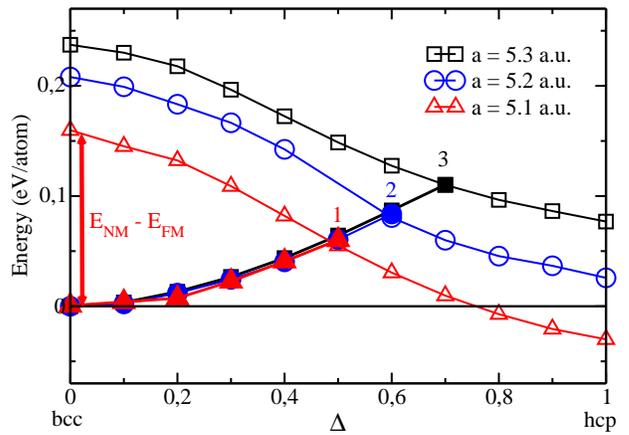}
\caption{\label{FIG_E_PATH} Energy as a function of the
parameter $\Delta$. Full symbols correspond to the FM state, open symbols - to
  the NM state. The vertical arrow indicates the energy difference $\Delta E = E_{NM}-E_{FM}$} 
\end{figure}

Fig. \ref{FIG_E_PATH} represents the total energy as a function of the
parameter $\Delta$, $E(\Delta)$, for different lattice parameters $a$ 
corresponding to different pressure values.   
The  $E(\Delta)$ curves have two minima corresponding to the FM bcc
structure ($\Delta 
= 0$) and NM hcp structure ($\Delta = 1$), which is a
quasi-equilibrium state. At low pressure the FM bcc structure of Fe is more
stable, while as pressure increases, the energy of the NM hcp structure
becomes deeper leading to the stability of this state. Note that the
$E(\Delta)$ curves (see Fig. \ref{FIG_E_PATH}) calculated for three different 
pressure values (or, equivalently, lattice parameters $a$ equal to
$5.1$, $5.2$, and $5.3$ a.u.) for the FM state of Fe, have nearly the
same dispersion. 
This is in line with previous phonon calculations exhibiting no
softening of the $N$-point phonon modes under pressure as discussed above. 

As follows
from Fig. \ref{FIG_ETOT}, the NM state of bcc Fe is higher in energy
than the FM state. $E_{FM}(\Delta)$ increases  
while $E_{NM}(\Delta)$ decreases monotonously with
$\Delta$ varying from 0 to 1 (open symbols in Fig. \ref{FIG_E_PATH}),
indicating the instability of NM bcc Fe with respect to this type of 
distortions, in line with the results on phonon calculations
\cite{ESEB98,HCG+02}.  
The 'critical' distortion values, $\Delta_c$, correspond to the
cross-points of the total energy curves  
as functions of the lattice distortion $\Delta$, calculated for the FM
($E_{FM}(\Delta)$) and NM ($E_{NM}(\Delta)$) states. I.e., at the
critical values of distortions $E_{FM}(\Delta_c) = E_{NM}(\Delta_c)$. 
The magnitude of the energy cusps  in Fig. \ref{FIG_E_PATH}, defined by
$E_{ci} = E_{NM} = E_{FM}$ for the various lattice parameters $a_i$, 
$E_{c1}$, $E_{c2}$, $E_{c3}$ and corresponding amplitude of the
'critical' distortions ($\Delta_{c1}$, $\Delta_{c2}$, and $\Delta_{c3}$)  
are getting smaller when the pressure increases.
%
%
\begin{figure}
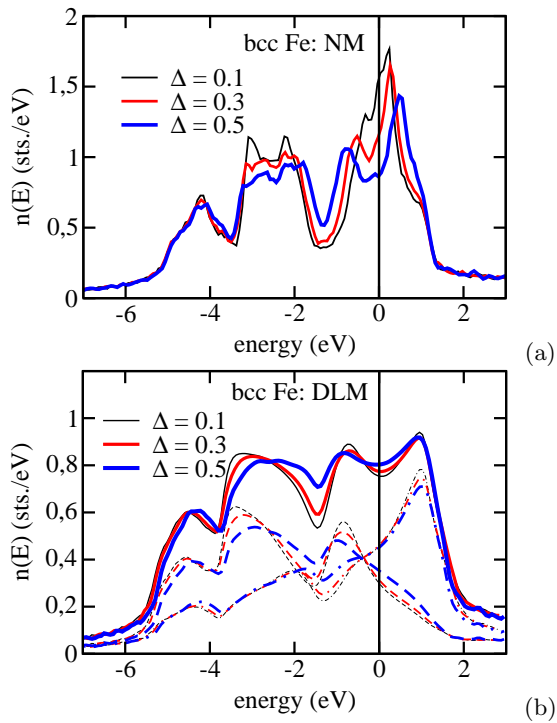

\includegraphics[width=0.37\textwidth,angle=0]{Figure_3_a.eps} \;(a)\;
\includegraphics[width=0.37\textwidth,angle=0]{Figure_3_b.eps} \;(b)
\caption{\label{FIG_DOS} Total DOS per atom as a functions of the
  parameter $\Delta$ for Pauli paramagnetic (non-magnetic) (a) and
  paramagnetic (via DLM calculations) (b) states of bcc Fe. Dashed (dashed-dotted) lines represent spin-up (spin-down) states.}
\end{figure}
%

Fig. \ref{FIG_DOS}(a) shows
the DOS, $n(E)$, corresponding to the NM state of distorted bcc Fe.
For small distortions,  $n(E)$  has a maximum at the Fermi level $E_F$
indicating the instability of the NM state. This DOS maximum
is formed by the double-degenerated $e_g$ electronic energy bands.
Therefore there are two
scenarios to remove the instability by breaking the symmetry of
the system: either due to structure distortion (e.g., Burgers
distortion) leading to the splitting of doubly degenerated  $e_g$ states 
at $E_F$ or due to spontaneous 
magnetization leading to the exchange splitting of the electronic states
having opposite spin directions.
At ambient pressure the second scenario is more favorable leading
to the stabilization of the bcc structure with FM order. 
Under high pressure, however, due to a reduced tendency towards
spontaneous magnetization via the Stoner mechanism the first
scenario (structure distortion) could be more preferable, as discussed
by Ekman \cite{ESEB98}. 
The same effect can be achieved even at lower pressure,
when the amplitude of the distortion is big enough, as demonstrated in
Fig. \ref{FIG_E_PATH} by the 'critical' points $1, 2$ and $3$. 
However, in this case the energy barrier $E_{FM}(\Delta_c) -
E_{FM}(\Delta = 0)$ is much too high to be overcome assuming the
fluctuation mechanism of phase transition, even at the pressure $p
\approx 23$~GPa ($a = 5.1$~a.u.), when the hcp structure becomes
energetically more preferable.   

So far we have discussed the stability of the system
with respect to lattice distortions assuming no deviations from perfect
FM order. 
To discuss in more details the interrelation between the structural
and magnetic degrees of freedom, the effect of spin moment fluctuations
should also be investigated. Moreover, such investigations are required
to account for the conditions of the experimental measurements performed
at room temperature.   
To take the effect of temperature induced spin fluctuations 
at $T > T_c$ into account, the paramagnetic state with non-zero local
moments of Fe was simulated using the so-called disordered local moment (DLM) 
scheme describing a random distribution of the local 
magnetic moments over the lattice sites. Accordingly, discussing the
results of calculations for the PM state we will use the term 'DLM
state' implying the approximation used in these calculations. 
The DLM calculations are done by means of the CPA alloy theory applied to the
effective alloy, Fe$_{x}^+$Fe$_{1-x}^-$, with equal amount of
'alloy' component ($x=0.5$) having the spin magnetic moment along  the $+{z}$
direction (Fe$^+$) and in the opposite direction (Fe$^-$)
\cite{SGP+84, GPS+85}. 

\begin{figure}
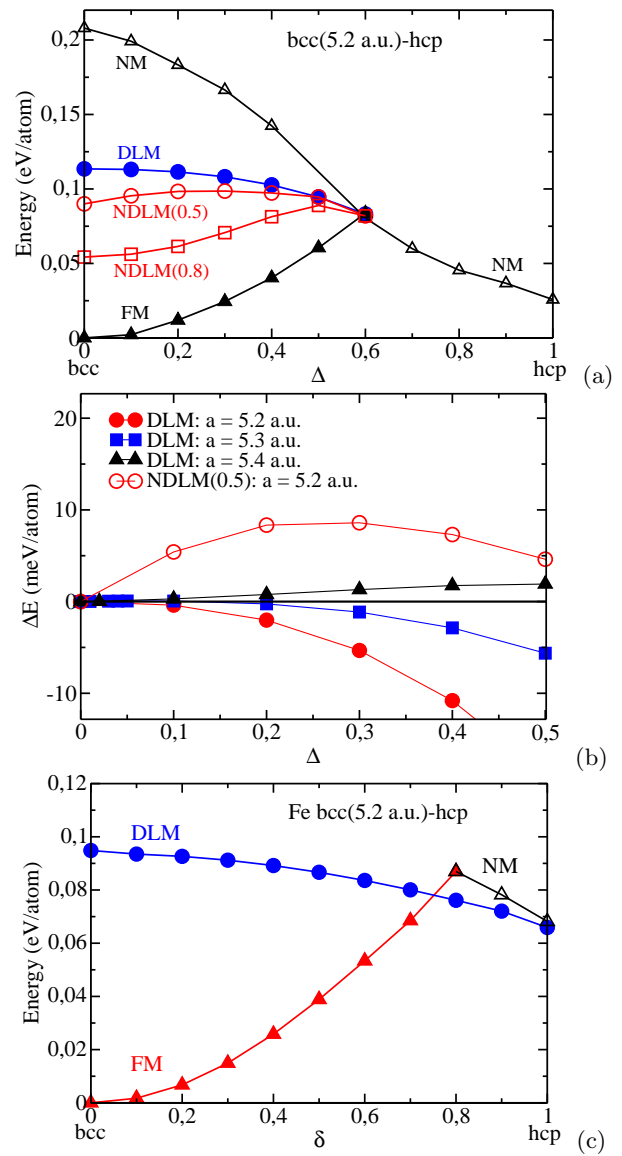

\includegraphics[width=0.41\textwidth,angle=0]{Figure_4_a.eps} \;(a)
\includegraphics[width=0.4\textwidth,angle=0]{Figure_4_b.eps} \;(b)
\includegraphics[width=0.4\textwidth,angle=0]{Figure_4_c.eps} \;(c)
\caption{\label{FIG_ETOT_DLM} Total energy as a function of the lattice
  distortion parameter $\Delta$ for bcc Fe with $a = 5.2$ a.u.: (a)
  comparison of the results for FM (triangles), PM (DLM, closed circles),
  partially disordered FM (NDLM) with non-compensated magnetic moment
  $M/M_s = 0.5$ (opened circles) and  $M/M_s = 0.8$ (opened squares);
(b) comparison of the DLM and NDLM ($M/M_s = 0.5$) results for different lattice
parameters; (c) total energy as a function of the amplitude of the $N$
point TA$_1$ phonon mode for FM state (closed triangles), NM state
(opened triangles) and DLM state (closed circles).
}
\end{figure}

The $E$ dependence on $\Delta$ for the PM state obtained (via DLM) 
at the pressure corresponding to $a = 5.2$ a.u.  
is shown in Fig. \ref{FIG_ETOT_DLM}(a) by solid 
circles. For $\Delta = 0$  (i.e. ideal bcc structure) the
difference $E_{PM}(\Delta = 0) - E_{FM}(\Delta =0)$ is about two times
smaller than $E_{NM}(\Delta = 0) - E_{FM}(\Delta =0)$.
$E_{PM}(\Delta)$ decreases slowly with $\Delta$ varying from 0 to 1,
indicating an instability of the DLM state of bcc Fe at this pressure value.
 Note that the $E(\Delta)$ dependence for DLM state, $E_{PM}(\Delta)$, is
 weaker than that corresponding to the NM state, $E_{NM}(\Delta)$. 
This behavior can be understood using the DOS plots shown in
Fig. \ref{FIG_DOS}. In the case of DLM state, one can clearly see a
local DOS minimum  (Fig. \ref{FIG_DOS}(b)) at the Fermi energy, created
by the exchange-split majority and minority $d$-states of Fe. The
dependence on $\Delta$ of the DOS obtained for DLM state is rather weak,
in contrast to the DOS obtained for the NM state
(Fig. \ref{FIG_DOS}(a)), having a pronounced maximum at $E_F$ strongly 
modified due to Burgers type of distortions. 

However, instability of the DLM state of bcc Fe with respect to the lattice distortion
$\Delta$ occurs only at the pressure exceeding a certain critical value
(closed circles and squares in Fig. \ref{FIG_ETOT_DLM}(b)). In the
vicinity of 
ambient pressure, the $E_{PM}(\Delta)$ curve has a minimum at $\Delta =
0$ (closed triangles in Fig. \ref{FIG_ETOT_DLM}(b)), that means
stability of this state with respect to Burgers distortions. 
 
To simulate the magnetic disorder corresponding to the temperature below the
critical one, $T_C$, so-called non-compensated DLM (NDLM) calculations have
been performed with the NDLM state simulated by an effective alloy
Fe$_{1-x}^+$Fe$_{x}^-$ with $x \in [0.0,0.5] $. 
In this case the normalized magnetic moment
$M/M_s = (M_sn_+ - M_sn_-)/M_s$ at each lattice site is equal to $(1-2x)$,
assuming $M_s$ to be a saturated local spin
magnetic moment of the Fe atoms. 
Open circles in Fig. \ref{FIG_ETOT_DLM}(a) correspond to the NDLM state of bcc
Fe ($a =5.2$ a.u.) with $M/M_s = 0.5$. In this case the curve
$E(\Delta)$ has a minimum at $\Delta = 0$ (open circles in
Fig. \ref{FIG_ETOT_DLM}(b)) and exhibits a slow increase with $\Delta$
increasing up to the critical displacement $\Delta_c \approx 0.25$.
The total energy 
$E_{NDLM}(\Delta = 0)$ decreases further, 
when $M/M_s$ changes up to $M/M_s = 1$, that is associated with the
temperature decrease and increase of FM order. As an example, open
squares in Fig. \ref{FIG_ETOT_DLM}(a) represent the $E_{NDLM}(\Delta)$
dependence for $M/M_s = 0.8$ (i.e., $x = 0.1$). 
Thus, the energy of Burgers type lattice distortions, $E(\Delta) - E(0)$,
decreases in the presence of temperature induced magnetic disorder
and close to the PM state (DLM state) becomes much smaller than in the case 
of perfect FM order. As a consequence, this leads to an increase of atomic displacements, caused directly by magnetic disorder
in the system.

Additional calculations have been performed to check explicitly the influence of
magnetic disorder on the energy of the $N$ point $TA_1$ phonon mode
(opposite displacement of adjacent (110) planes along the $[110]_{bcc}$
direction), i.e., without accounting for shear deformation along the [001] direction
present in the case of the Burgers type deformation. 
The results are shown in Fig. \ref{FIG_ETOT_DLM}(c). 
The energy dependence $E(\delta)$ on the atomic displacements
 $\delta$ is rather similar to that obtained for the Burgers deformations,
 shown in Fig. \ref{FIG_ETOT_DLM}(a).
Thus, these results show an explicit evidence of the crucial role
of magnetic disorder for the softening of the $N$ point $TA_1$ phonon  mode
making it responsible as the driving mechanism for the bcc to hcp transition.

\begin{figure}
 \includegraphics[width=0.35\textwidth,angle=0]{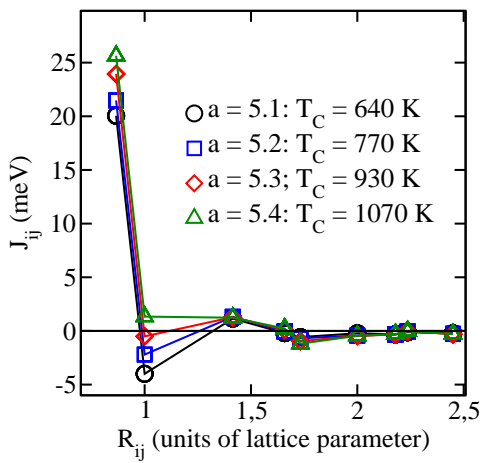}
 \caption{\label{FIG_JXC_TVIB} (a) Exchange coupling parameters for bcc Fe at
   different lattice parameter via DLM calculations for case (1) (see text). }
 \end{figure}

\begin{figure}
\includegraphics[width=0.45\textwidth,clip]{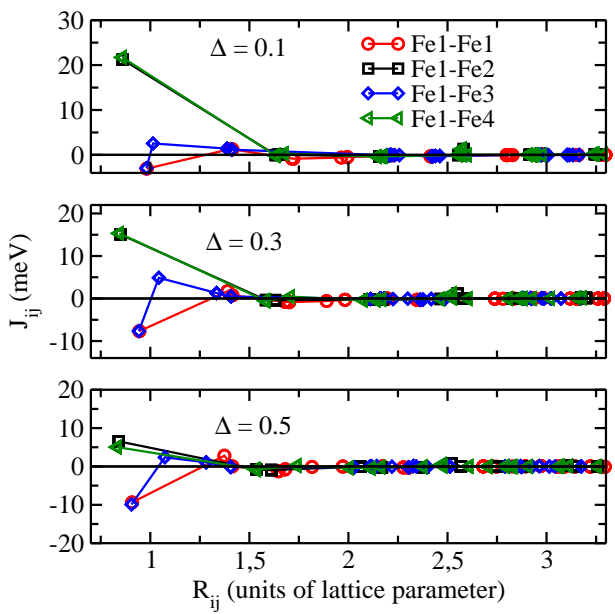}
\caption{\label{FIG_JIJ_DELTA} Dependence of the exchange coupling
  parameters on the amplitude of lattice distortion for bcc Fe under
  pressure corresponding to lattice parameter $a = 5.2$~a.u. for case
  (2) (see text)} 
\end{figure}

Additional investigations have been performed to show an effect of
lattice displacements on the magnetic order in the system.
Considering atomic fluctuations corresponding to the Burgers type of
lattice  deformation \cite{FS08}, 
two series of the exchange coupling parameters, $J_{ij}$,
have been calculated for
the DLM state: (1) for different pressure values
(i.e. different lattice parameters) for the perfect bcc structure and (2)
for the distorted bcc lattice with $a = 5.2$ a.u. with different 
distortion parameter $\Delta $.  
Fig. \ref{FIG_JXC_TVIB} shows the results for the case (1),
for which  $J_{ij}$ values are presented together with corresponding 
Curie temperatures, determined by means of MC simulation, 
and exhibiting their decrease when the pressure increases.
The corresponding exchange coupling parameters $J_{ij}$ for distorted
bcc Fe (case (2)) are shown in Fig. \ref{FIG_JIJ_DELTA} where 
one can easily see that lattice Burgers distortions are accompanied 
by pronounced variations of the exchange coupling parameters.
In particular, we can point out a decrease of the FM and increase of the  
AFM exchange interactions when the $\Delta$ parameter increases.

\begin{figure}
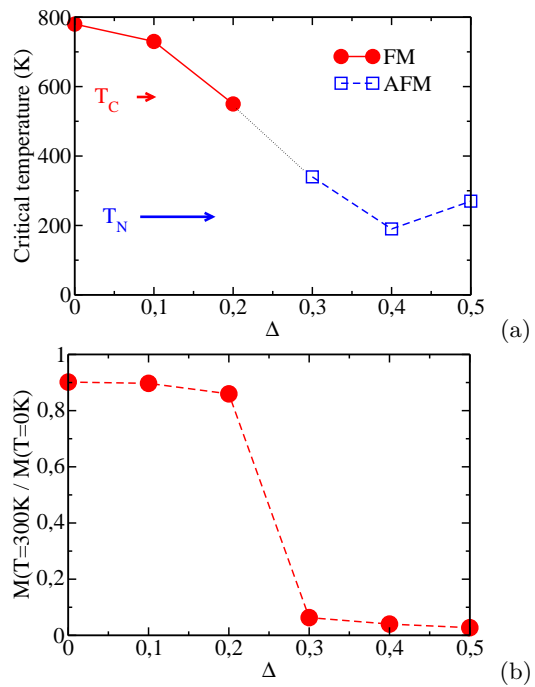

\includegraphics[width=0.35\textwidth,angle=0]{Figure_7_a.eps} \;(a) 
\includegraphics[width=0.35\textwidth,angle=0]{Figure_7_b.eps} \;(b) 
\caption{\label{FIG_MC} Results of MC simulations: (a) Critical
  temperature as a function of lattice distortion $\Delta$ for bcc Fe
  with $a = 5.2$  a.u.; (b) the normalized average magnetic moment
  corresponding to the temperature $T = 300$ K.} 
\end{figure}

Assuming that the biggest amplitude of atomic displacements is related
to this type of distortions (due to their softening),
the corresponding critical temperature have been calculated
using the simulations for different values for the $\Delta$ parameter
(Fig. \ref{FIG_MC}(a)). This simplified approach was used to demonstrate 
the effect of atomic displacements on the finite-temperature magnetic
properties of Fe under fixed pressure (corresponding to $a = 5.2$
a.u.). At small distortions the system has FM order with 
the Curie temperature decreasing upon $\Delta$ increase.
At  $\Delta \geq 0.3$ the system exhibit AFM properties (non-collinear
or collinear, depending on $\Delta$). 
The average magnetic moment calculated for different $\Delta$ values
at the temperature $T = 300$ K is presented in Fig. \ref{FIG_MC}(b).
 It drops down at rather small lattice distortions, $\Delta > 0.2$,
first, due to transition to the state with non-collinear magnetic
structure and then, for $\Delta > 0.3$, to the PM state.

In spite the
simplifications used for our analysis, the results presented above
directly demonstrate the strong mutual influence of
the lattice Burgers distortion and spin moment 
fluctuations resulting in a pronounced pressure and temperature dependence 
of the geometric and magnetic structure in the system.  
Both effects are the counterparts of the mechanism 
leading to a softening of the corresponding phonon modes responsible for
the instability of the bcc state of Fe and leading to the bcc-hcp transition.

\subsection*{II.D. Experimental observations vs theory: \\
              K-edge XMCD}

Experimental investigations on bcc Fe under pressure
have been performed using X-ray absorption spectroscopy at the K-edge of
Fe, with the XANES and XMCD spectra measured simultaneously
\cite{MBI+04}. This allowed 
to check experimentally the role of magnetic order for the stability of
bcc-Fe, as discussed in the literature (see, e.g. \cite{ESEB98}).  
In particular, a synchronous decrease of the 'structural' and 'magnetic'
XAS signals related to the ferromagnetic bcc phase of Fe
would be expected at the critical pressure, where this loss
of stability is associated with a transition to the Pauli paramagnetic
(i.e., NM) state.    

\begin{figure}
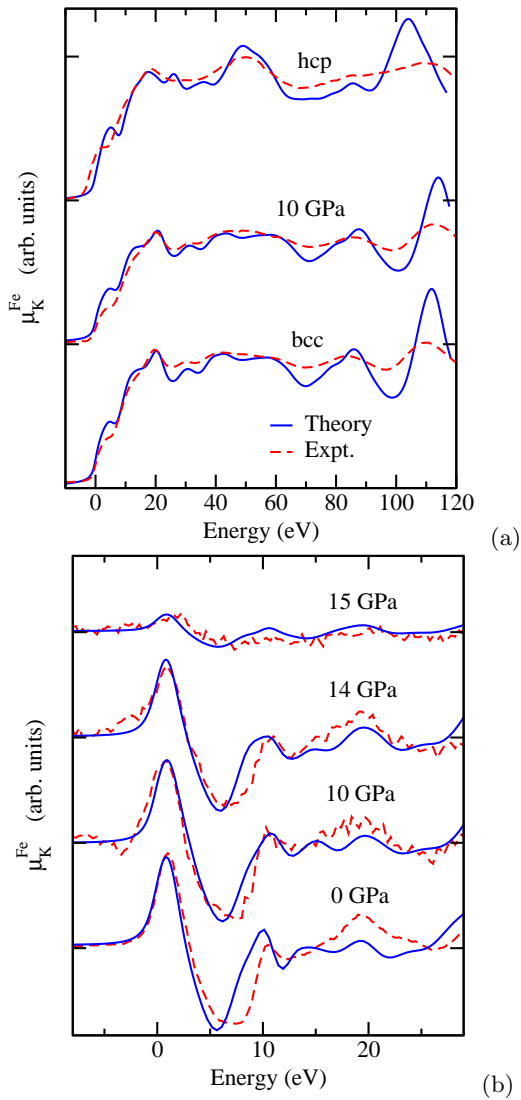

\includegraphics[height=0.40\textwidth,angle=0]{Figure_8_a.eps} \;(a)\;
\includegraphics[height=0.4\textwidth,angle=0]{Figure_8_b.eps} \;(b)
\caption{\label{FIG_XAS} K-edge XANES for bcc Fe at different pressures
  as well as hcp Fe at a pressure above the critical one: (a) XANES and
  (b) XMCD spectra. Solid and dashed lines correspond to theoretical
 and experimental results, respectively.} 
\end{figure}

Theoretical calculations of the X-ray
absorption and XMCD spectra at the K-edge of Fe have been performed for
different lattice parameters corresponding to the pressure values below
the critical one. These results are compared in Fig. \ref{FIG_XAS} with
experimental spectra 
measured at different pressures below the critical value. 
In this case the spectra are associated to the bcc
structure with a slow variation upon pressure increase caused by the
pressure induced variation of the lattice parameter. As one can see,
the theoretical calculations reproduce the experimental XANES results quite
well. The same applies for the XMCD spectra. 
At the pressure above the critical value, the XANES spectrum reflects
the hcp structure  
and is again in good agreement with experiment (see Fig. \ref{FIG_XAS}).
In this case the XMCD signal is very weak and corresponds to the remnant
bcc phase at this pressure, which disappears completely if the pressure
in further increased.

To explain the pressure dependence of the experimental XMCD
spectra, we refer to the theoretical results discussed above.
The XANES experiment performed at the $K$ edge of Fe 
implies that the induced orbital polarization of the $p$-electrons is
probed. In ordered FM systems this implies
that the K-edge XMCD signal should be roughly proportional to the spin
magnetic moment of the 3d electrons (see, e.g., \cite{IKY+07, TKP+11}
and references therein).  
At finite temperature, however, a decrease of the dichroic signal can
occur due to an increase of the magnetic disorder in the system, even 
for weak changes of the local magnetic moment.
In the present case of bcc Fe at fixed (room) temperature, the increase
of the magnetic disorder is governed by the increasing pressure.
At the same time, the XANES signal does not change because
the bcc structure remains stable until the critical value of the
magnetic disorder (associated with a critical pressure) is reached. This
is indeed observed in the experimental XMCD spectra. A further pressure
increase should result in a quick drop down for the average magnetic
moment leading to the instability of the bcc structure and to a
transition to the hcp phase, as it is observed in the XANES and XMCD
spectra.

\section*{III. Summary}

In summary, our investigations led to the following results:
(i) Assuming no structural nor spin distortions for bcc Fe,
a transition to the hcp structure can occur only at very high pressures,
corresponding to a lattice parameter $\approx 4.6$ a.u., 
at which the system becomes paramagnetic;
(ii) Burgers-type lattice distortions in collinear FM bcc Fe can lead to 
a transition to the NM state and as a result to the  
bcc-hcp transition at the pressure close to the one observed
experimentally, implying fluctuation mechanism of transition.
However, the energy of lattice fluctuations required for the FM-PM
transition is too high; 
(iii) DLM calculations show an instability of spin disordered
bcc Fe upon a pressure increase. This requires the transition to
the PM state. To get this condition at the temperature of measurements,
$T_e$, the Curie temperature should be low enough, i.e. $T_C < T_e$.
However, this is not the case. On the other hand,
even a partial spin disorder can result in the decrease of the energy of
lattice fluctuations required for transition to the PM state.
This implies that both effects, spin and lattice fluctuations,  
are the counterparts of the mechanism leading to a softening of the
corresponding phonon modes responsible for the instability of the bcc
state of Fe and leading to bcc-hcp transition.

\section{Acknowledgements}

Financial support within the framework of Deutsches
Elektronen-Synchrotron DESY (BMBF) 05K13WMA program
and the priority Schwerpunktprogramm SPP 1415
is gratefully acknowledged.

\section*{Appendix}

\begin{figure}
\begin{center}
\vspace{-0.2cm}
\includegraphics[width=0.11\textwidth,angle=0]{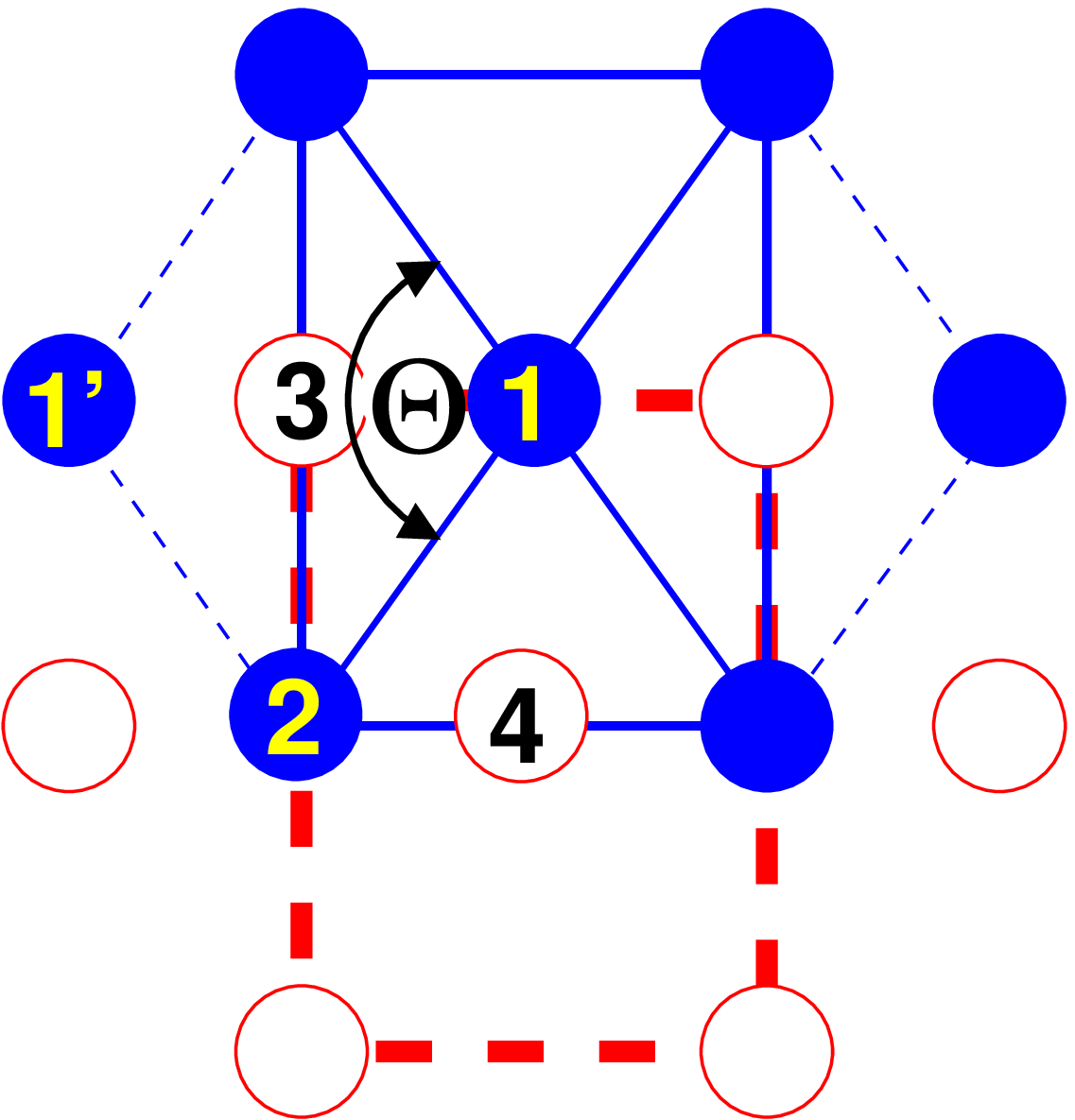}\;\;(a)
\includegraphics[width=0.11\textwidth,angle=0]{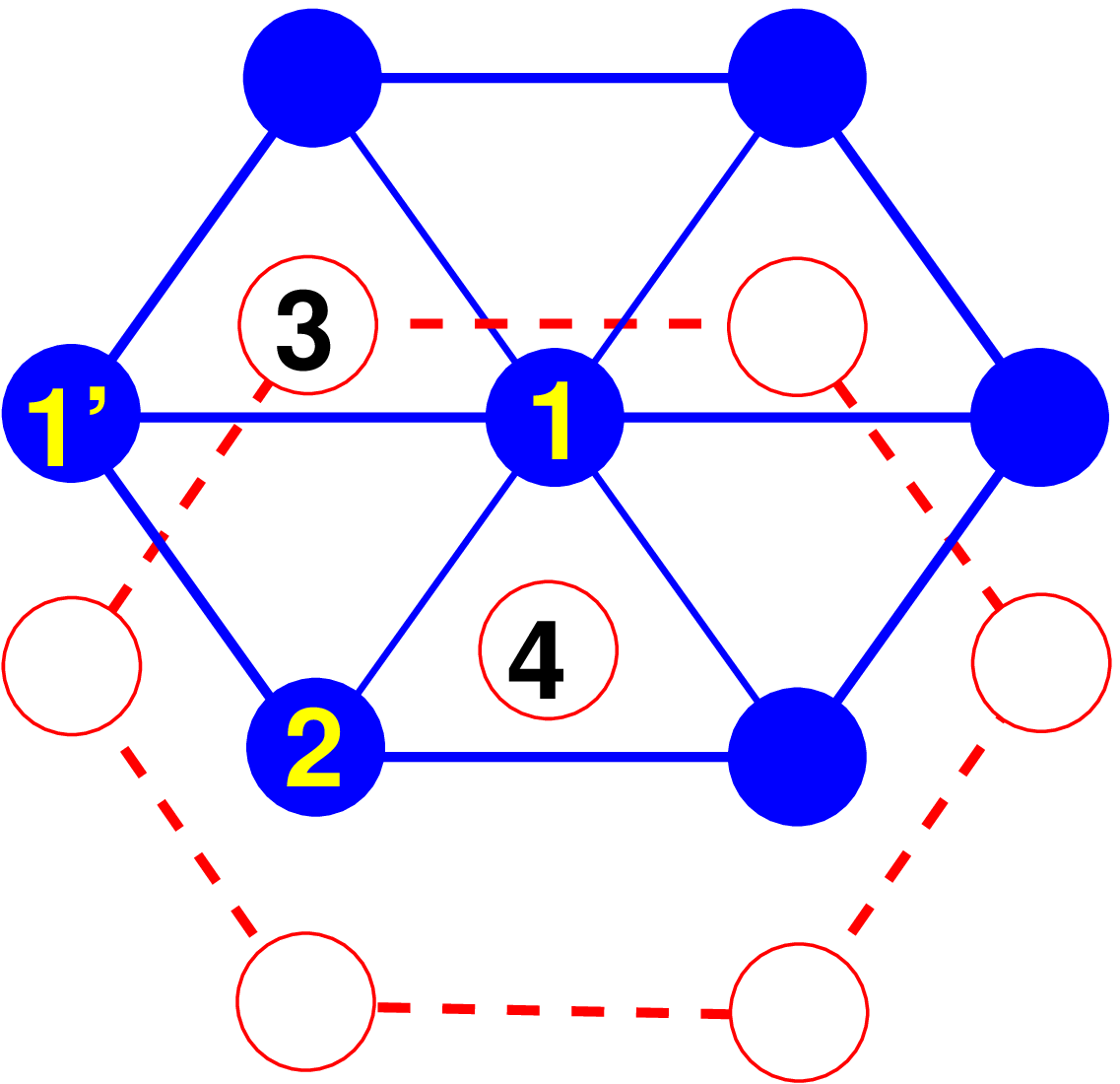}\;\;(b)
\includegraphics[width=0.11\textwidth,angle=0]{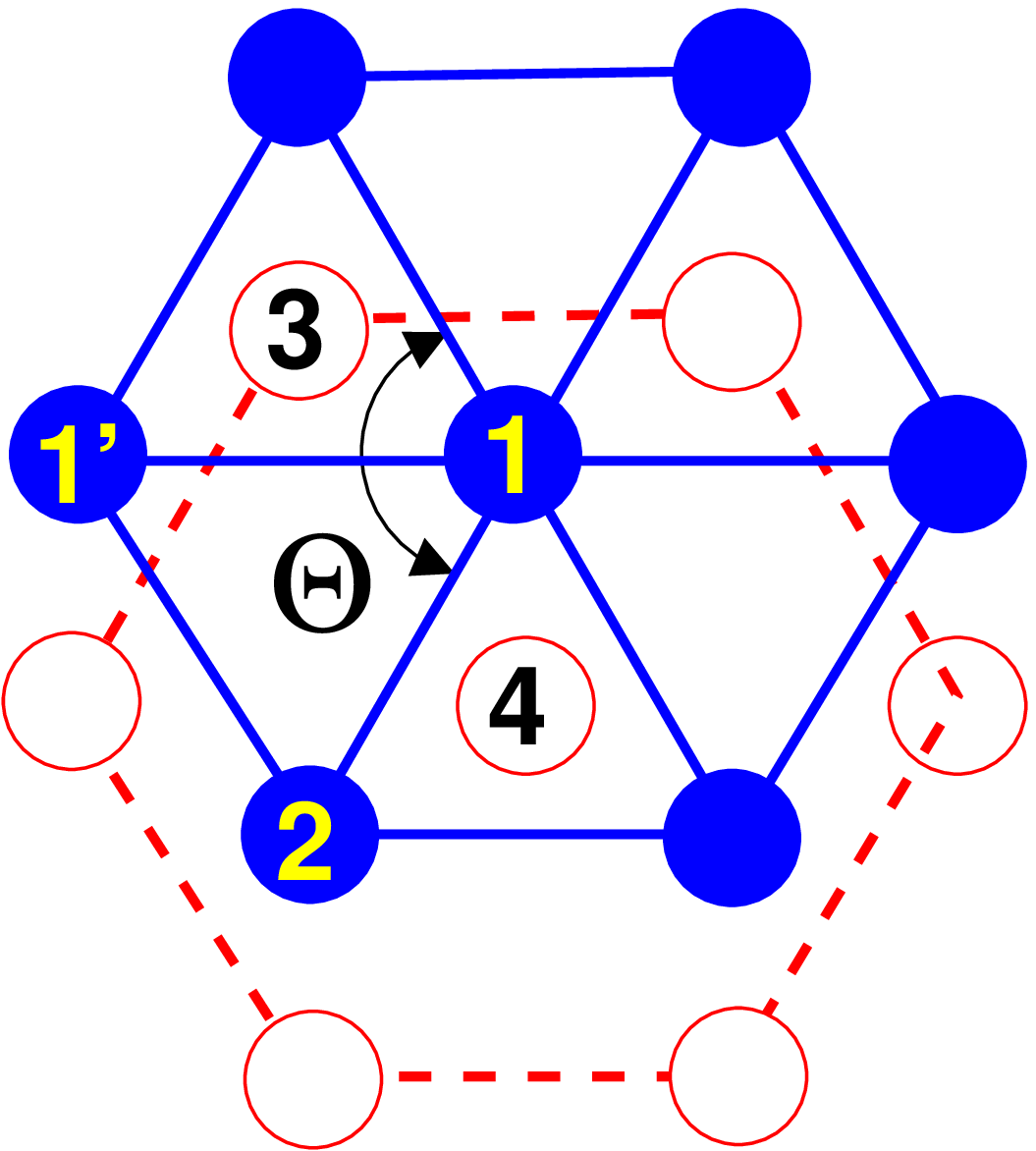}\;\;(c)
\caption{\label{FIG_Burgers} Transformation from bcc to hcp structure
  according to Burgers scheme: (a) (110) planes of the bcc structure; (b)
  opposite displacement of adjacent (110) planes along the $[110]_{bcc}$
  direction; (c) hcp structure - after shear deformation along [001]
  direction with $\theta$ changing from $109.5^o$ in the case of bcc
  structure to $120^o$.
 }
\end{center}
\end{figure}

According to Friak and Sob \cite{FS08}, transformations using only
one parameter $\Delta$ includes both transformations represented by 
the parameters $\delta$ or $\theta$ (see Fig. \ref{FIG_Burgers}),
avoiding the high-energy configurations occurring upon their independent
variation. Variation of $\Delta$ parameter keeps the volume
$V(\Delta)$ unchanged and leads to the hcp structure with the ratio $c/a
= \sqrt{8/3}$, following the path being close to the minimum energy
path \cite{FS08}. The lattice parameters for the distorted  orthorombic
structure with 4 atoms/u.c. varies as follows:
\begin{eqnarray*}
a &=& a_0\sqrt{2}/(V/V_0)^{1/3} \\
b &=& a[\Delta(2\sqrt{3} - 3\sqrt{2})/6 + 2\sqrt{2}] \\
c &=& a[\Delta(2\sqrt{2} - 3)/3 + 1] \\
V/V_0 &=& \sqrt{2}[\Delta(2\sqrt{3} - 3\sqrt{2})/6 +
2\sqrt{2}][\Delta(2\sqrt{3} - 3)/3 + 1] 
\end{eqnarray*}
The atomic positions in the unit cell with distortion are  $(0,0,0),
(1/2,1/2,0), (1/2-\Delta/6,0,1/2), (-\Delta/6,1/2,1/2)$.   

\end{document}